\documentclass[journal]{IEEEtran}

\ifCLASSINFOpdf
\else
\fi

\usepackage[utf8]{inputenc}
\usepackage{cite}
\usepackage{graphicx}
\usepackage{amssymb}
\usepackage{amsmath}
\usepackage{booktabs}
\usepackage{array}
\usepackage{subcaption}
\usepackage{amsfonts, amsthm}
\usepackage{wrapfig} 
\usepackage{cite}

\usepackage{bbm}
\usepackage{lipsum}
\usepackage{changepage}
\usepackage{color,soul}
\usepackage{enumitem}   
\usepackage{relsize}

\makeatletter
\newcommand{\thickhline}{%
    \noalign {\ifnum 0=`}\fi \hrule height 1pt
    \futurelet \reserved@a \@xhline
}
\newcolumntype{"}{@{\hskip\tabcolsep\vrule width 1pt\hskip\tabcolsep}}
\makeatother

\usepackage{framed,multirow}
\usepackage{latexsym}
\usepackage{url}
\usepackage{xcolor}
\usepackage{bm}

\usepackage{todonotes}
\newcolumntype{L}[1]{>{\raggedright\let\newline\\\arraybackslash\hspace{0pt}}m{#1}}
\newcolumntype{C}[1]{>{\centering\let\newline\\\arraybackslash\hspace{0pt}}m{#1}}
\newcolumntype{R}[1]{>{\raggedleft\let\newline\\\arraybackslash\hspace{0pt}}m{#1}}

\usepackage{lipsum}
\usepackage{changepage}
\usepackage{color,soul}
\usepackage{enumitem}   
\usepackage{float}

\usepackage{booktabs,xcolor}
\definecolor{lightgray}{gray}{0.9}

\usepackage{tikz}

\let\OLDthebibliography\thebibliography
\renewcommand\thebibliography[1]{
  \OLDthebibliography{#1}
  \setlength{\parskip}{0pt}
  \setlength{\itemsep}{0pt plus 0.3ex}
}

\usepackage{hyperref}

\usepackage{afterpage}

\newcolumntype{Y}{>{\centering\arraybackslash}X}

\begin{document}

\title{Detailed delineation of the fetal brain in diffusion MRI via multi-task learning}

\author{Davood Karimi$^1$, Camilo Calixto$^{1,2}$, Haykel Snoussi$^1$, Maria Camila Cortes-Albornoz$^3$, \\ Clemente Velasco-Annis$^1$, Caitlin Rollins$^1$, Camilo Jaimes$^3$, Ali Gholipour$^{1,4}$, Simon K. Warfield$^1$  \\ $^1$ Boston Children's Hospital and Harvard Medical School, Boston, MA \\ $^2$ Elmhurst Hospital Center and Icahn School of Medicine at Mount Sinai, New York, NY \\ $^3$ Massachusetts General Hospital and Harvard Medical School, Boston, MA \\ $^4$ Department of Radiological Sciences, University of California Irvine, Irvine, CA }

\maketitle

\begin{abstract}

Diffusion-weighted MRI is increasingly used to study the normal and abnormal development of fetal brain in-utero. Recent studies have shown that dMRI can offer invaluable insights into the neurodevelopmental processes in the fetal stage. However, because of the low data quality and rapid brain development, reliable analysis of fetal dMRI data requires dedicated computational methods that are currently unavailable. The lack of automated methods for fast, accurate, and reproducible data analysis has seriously limited our ability to tap the potential of fetal brain dMRI for medical and scientific applications. In this work, we developed and validated a unified computational framework to (1) segment the brain tissue into white matter, cortical/subcortical gray matter, and cerebrospinal fluid, (2) segment 31 distinct white matter tracts, and (3) parcellate the brain's cortex and delineate the deep gray nuclei and white matter structures into 96 anatomically meaningful regions. We utilized a set of manual, semi-automatic, and automatic approaches to annotate 97 fetal brains. Using these labels, we developed and validated a multi-task deep learning method to perform the three computations. Our evaluations show that the new method can accurately carry out all three tasks, achieving a mean Dice similarity coefficient of 0.865 on tissue segmentation, 0.825 on white matter tract segmentation, and 0.819 on parcellation. The proposed method can greatly advance the field of fetal neuroimaging as it can lead to substantial improvements in fetal brain tractography, tract-specific analysis, and structural connectivity assessment.

\end{abstract}


\section{Introduction}
\label{sec:introduction}

\subsection{Background and motivation}
\label{sec:background}

The fetal period represents the most dynamic stage in brain development, when a series of complex processes work together to transform a simple neural tube into a highly structured adult-like brain within a few months \cite{kostovic2006development}. During this period, the fetal brain is particularly vulnerable to diseases and environmental factors that can result in lifelong neurodevelopmental and psychiatric disorders \cite{donofrio2011impact, lynch2009epidemiology}. Therefore, accurate and detailed assessment of the fetal brain is crucial in neuroscience and medicine. Medical imaging has played an important role in this regard. Magnetic resonance imaging (MRI), in particular, has provided a wealth of information about brain development in utero. Fetal MRI can also serve as an indispensable complement to ultrasound imaging for clinical fetal brain assessment \cite{egana2013differences, gholipour2014fetal}.

Diffusion-weighted MRI (dMRI), in particular, offers unique capabilities for studying the brain. It enables mapping of the white matter microstructure and quantitative structural connectivity analysis \cite{ouyang2019delineation, huang2009anatomical}. Fetal dMRI has witnessed an accelerating progress in its ability to reveal important information about the development and maturation of the brain white matter in utero \cite{khan2019fetal, calixto2024detailed}. However, the potential of dMRI to probe the fetal brain still remains largely untapped \cite{dubois2014early, jakab2017utero}. This is in part due to the technical challenges of analyzing fetal brain dMRI data. Compared with adult brain imaging, fetal dMRI data has lower signal-to-noise ratio. Fetal motion during image acquisition can be large and unpredictable. Another important issue is the rapid development of brain microstructure and macrostructure, which makes it difficult to develop methods that can work reliably across the gestational age.

The overwhelming majority of existing dMRI data analysis methods and software have been developed for postnatal and adult brains. Because of the vast differences between fetal and adult brains, these tools are not suited for fetal data. Reliable analysis of fetal data requires dedicated computational methods, which are mostly lacking. In particular, although deep learning has led to significant breakthroughs in dMRI analysis \cite{karimi2024diffusion}, very little work has been done to address the needs of fetal dMRI. A major reason for this gap in technology is paucity of training data. Collecting and labeling fetal MRI data is costly and challenging \cite{payette2023fetal, calixto2023detailed}. This problem is further complicated by the rapid and dramatic changes in the brain size, shape, and structure in utero. In order to develop and validate accurate machine learning methods, an adequate number of images need to be labeled at every gestational age to capture the inter-subject heterogeneity and the temporal variability due to brain development \cite{payette2023fetal, karimi2023learning}.

The goal of this work is to develop a method for automatic delineation of fetal brain structures in dMRI. We adopt a multi-task learning approach and develop a single model to delineate the fetal brain in three distinct ways:

\begin{enumerate}

\item Tissue segmentation: This involves segmenting the brain into four tissue classes: white matter (WM), cortical gray matter (CGM), sub-cortical gray matter (SGM), and cerebrospinal fluid (CSF).

\item White matter tract segmentation: Under this scheme, a selection of 31 WM tracts are delineated. The tract names and abbreviations used in this paper are presented in Table \ref{table:tract_names} in the Supplementary Materials.

\item Parcellation: This scheme involves parcellation of the brain cortex into anatomically meaningful regions and delineation of deep gray nuclei and white matter structures. It includes 96 labels, including 47 with bilateral representation and 2 mid-line structures. The full names and abbreviations for these regions are provided in Table \ref{table:parcel_names} in the Supplementary Materials.

\end{enumerate}

Each of these three schemes has important practical applications. Brain tissue segmentation is crucial for quantifying normal and abnormal brain growth, as well as for various tractography methods, such as anatomically-constrained tractography \cite{smith2012anatomically}, surface-enhanced tractography \cite{st2018surface}, and machine learning-based tractography \cite{poulin2019tractography}. Identification and delineation of individual white matter tracts can enable tract-specific analysis for studying microstructural and macrostructural alterations in specific tracts due to normal brain development or pathologies. Parcellation of the cortex and segmentation of the deep gray nuclei and white matter is essential for defining connectome nodes for structural connectivity assessment, where accurate and reproducible parcellation is crucial \cite{rheault2020common}. Additionally,  this may also be used in post-processing of tractography data \cite{wassermann2016white}. Therefore, the proposed method can advance the field of fetal brain imaging in multiple important ways.

\subsection{Related works}
\label{sec:related_works}

\subsubsection{Automatic delineation of the fetal brain in MRI and dMRI}
\label{sec:fetal_brain_segmentation}

Most prior works on segmentation of fetal brain images have focused on structural MRI \cite{makropoulos2018review}. Segmentation of the fetal cortical gray matter (only one class label) has been attempted in several works \cite{dou2020deep, dumast2021segmentation}. Several studies have segmented additional structures such as the white matter, CSF, and subcortical structures \cite{khalili2019automatic, payette2020efficient}. There have also been limited efforts to parcellate the fetal brain gray matter in T2-weighted MRI \cite{you2024automatic}. In terms of methodology, recent comparisons have shown that deep learning techniques outperform the more classical approaches, such as multi-atlas segmentation \cite{payette2021automatic, karimi2023learning}. However, none of these works have addressed the segmentation of fetal brain tissue directly in the dMRI space. Similarly, we are unaware of any prior work for automatic delineation of fetal brain white matter tracts or parcellation of gray matter in dMRI.

Low tissue contrast and the rapid pace of development of neuronal structures make it difficult to accurately annotate fetal brain images. Some prior studies have devised innovative methods to tackle this problem. For cortical gray matter segmentation, one study generated noisy annotations using an automatic segmentation technique that had been originally devised for neonatal brains \cite{fetit2020deep}. For tissue segmentation, another study used a multi-atlas method to label 272 fetal brain images, estimated the boundary uncertainty of each structure, and proposed a label smoothing technique to account for noisy training labels \cite{karimi2023learning}. In general, obtaining adequate training data is a persistent challenge for developing deep learning methods to delineate structures of interest in fetal brain images.

\subsubsection{Multi-task learning}
\label{sec:multitask_learning}

Multi-task learning (MTL) is a widely used approach in machine learning \cite{caruana1997multitask}. It refers to approaches that jointly address multiple machine learning tasks by leveraging the similarities and/or differences between the tasks. It is anticipated that a well-designed MTL approach should lead to better prediction accuracy than when training separate models for each task \cite{caruana1997multitask, crawshaw2020multi, vandenhende2021multi}. 

Multi-task learning has also become a popular approach in training deep neural networks \cite{zhang2014facial, misra2016cross, dai2016instance}. By addressing multiple tasks in a single framework, MTL approaches in deep learning have the potential to improve data efficiency and reduce the risk of overfitting. Numerous studies have shown that MTL can lead to improved task performance. For medical image analysis applications such as segmentation, too, prior studies have demonstrated the effectiveness of MTL. It has been shown that training a single model to segment different organs in different imaging modalities can result in higher segmentation performance compared with separate models trained to perform individual segmentation tasks \cite{moeskops2016deep, karimi2022improving}. A review of MTL methods in deep learning can be found in \cite{crawshaw2020multi}. Studies that are most related to the present work are multi-task \emph{dense prediction} techniques, which have been surveyed in \cite{vandenhende2021multi}.

Along with their benefits, MTL approaches pose new challenges. Different tasks may have conflicting needs, which can reduce the performance on some of the tasks involved. This situation is often referred to as ``negative transfer'' and is very common in MTL \cite{liu2019loss}. Much research has been devoted to automatically identifying task groupings that can optimally benefit from MTL \cite{fifty2021efficiently}. Proper balancing of the loss functions for different tasks can be very challenging. Various methods based on task uncertainty, learning speed, and task performance have been proposed to systematically balance different task objectives \cite{kendall2018multi, liu2019end, jean2019adaptive}. Other factors that can significantly influence the effectiveness of MTL include model architecture and optimization procedures. Effective sharing of parameters/activations among the tasks \cite{vandenhende2019branched, duong2015low}, manipulation of the optimization gradients \cite{chaudhry2018efficient}, and task scheduling and prioritization \cite{bengio2009curriculum, guo2018dynamic} are some of the techniques that are widely used to enable MTL in challenging applications.

\subsection{Contributions of this work}
\label{sec:contributions}

In this work, we developed and validated a unified framework to delineate the fetal brain in the dMRI space in three ways: (1) tissue segmentation, (2) white matter tract segmentation, and (3) parcellation. The deep learning methodology that is employed in this work is not entirely novel; rather, we have developed a novel computational framework that builds upon existing techniques. Nonetheless, our work is the first attempt to delineate the fetal brain directly in dMRI. To the best of our knowledge, no existing method offers any of the three capabilities that are enabled by our proposed method.

In order to generate the training data, we annotated a total of 97 fetal brain images using a combination of manual labeling by human experts, semi-automatic methods, and automatic techniques. The proposed method consists of a multi-task deep learning model that leverages convolutional and attention mechanisms. Our quantitative evaluations and human expert assessments show that the proposed method can achieve highly accurate results on all three tasks. In addition to sharing our source code for model training, we release our trained model as a Docker image to facilitate its use and integration into fetal dMRI pipelines by other investigators.

\section{Methods}
\label{sec:methods}

\subsection{Image data acquisition and preprocessing}
\label{sec:data_imaging}

The fetal brain MRI data used in this work were acquired at Boston Children’s Hospital. The study was approved by the institutional review board, and written informed consent was obtained from all participants. A total of 97 fetuses between 23 and 36 weeks of gestational age (GA) were included in this study. The dMRI scans consisted of single-shell measurements at b=500 that were acquired along orthogonal planes with respect to the fetal head. Isotropic dMRI volumes at a voxel size of 1.2 mm were reconstructed for each fetus using a slice-to-volume registration method \cite{marami2017temporal}. Subsequently, the diffusion tensor image (DTI) was computed using a weighted linear least squares method. A total of 28 of the subjects were held out for final model validation. The remaining 69 subjects were used for model development.

\subsection{Annotation procedures}
\label{sec:annotation}

A major component of our efforts in this work included image annotations to generate the required labels. Given the highly different nature of the three tasks considered in this work, we followed three different approaches to generate the labels for each task.

\subsubsection{Annotation for tissue segmentation}
\label{sec:annotation_segmentation}

We generated segmentation labels on the training subjects using a semi-automatic method. In our previous works, we created a DTI atlas of the fetal brain and manually segmented this atlas \cite{calixto2023detailed}. This atlas consists of 12 labels for fetuses below 31 gestational weeks and 11 labels for fetuses at 31 gestational weeks and above. The additional label in younger brains is because of the presence of two transient zones (i.e., the subplate zone and the intermediate zone).

To segment each training image in this work, we registered the three atlases closest in gestational age to the fetus. For example, if the fetus was at 30 gestational weeks,  we registered atlases at weeks 29, 30, and 31 To ensure accurate alignment, we employed a deformable diffusion tensor-based registration method \cite{zhang2006deformable}. The computed registration transforms were applied to the atlas segmentation maps to align them to the fetal subject. Subsequently, the probabilistic STAPLE algorithm \cite{akhondi2013simultaneous} was applied to fuse the registered labels and arrive at one segmentation map for the fetus. Then, two research fellows with medical training performed manual refinements if necessary, and a board-certified neuroradiologist with fellowship training in pediatric neuroradiology reviewed the final segmentation maps and revised them as needed. The atlas labels \cite{calixto2023detailed} were merged to create four tissue segmentation labels: WM, CGM, SGM, and CSF.

Out of the 28 test subjects, 17 were segmented using the same approach as the training images (as described above). For the remaining 11 test subjects, two experts manually generated the tissue segmentation labels. These manual segmentations were also reviewed and revised by two neuroradiologists with fellowship training in pediatric neuroradiology. This ensured optimal ground truth labels for validation of the new method. Due to limitations in resources, we couldn't manually segment a larger number of test subjects as it required approximately one week of an expert's time for each image, in addition to another expert review and revision.

\subsubsection{Annotation of white matter tracts}
\label{sec:annotation_tracts}

We generated labels depicting a set of 31 white matter tracts by following the steps described below.

\begin{enumerate}

\item We computed whole-brain tractograms using the iFOD2 tractography algorithm \cite{tournier2010improved}. Given the highly varying curvatures of different tracts, we reconstructed three tractograms using three different tractography angle thresholds of $15^{\circ}$, $20^{\circ}$, and $25^{\circ}$. This ensured that, in most cases, each tract was fully reconstructed with at least one of the angle thresholds. Each tractogram consisted of 5 million streamlines.

\item An automatic method \cite{wassermann2016white} was applied to extract streamline bundles representing a total of 55 distinct tracts from each of the three tractograms for each fetus.

\item An expert examined the three streamline bundles for every fetus/tract and selected the one that most accurately and most fully covered the anatomical tract of interest. If none of the reconstructions was acceptable, it was marked as such by the expert.

\item Each tract's streamline representation was converted to a binary mask by computing the streamline density map and removing the voxels where streamline density was less than the 5\textsuperscript{th} percentile of non-zero density values. We merged bilateral tract pairs (e.g., left and right IFO) into one label, reducing the total number of tract labels to be segmented from 55 to 31. The tract names are listed in Table \ref{table:tract_names} in the Supplementary Materials.

\end{enumerate}

\subsubsection{Annotation for parcellation and delineation of the deep gray nuclei and white matter structures}
\label{sec:annotation_parcellation}

The segmentation process was conducted in two main steps. First, we utilized the initial segmentation maps obtained during the tissue segmentation phase (after expert refinement but before merging into tissue classes) to delineate all brain structures, including cortical gray matter, deep gray nuclei, and white matter structures. Subsequently, we parcellated cortex labels (see below), which were specifically assigned to the CGM and periventricular white matter regions. For the test subjects, these parcellated cortex labels were applied to either manual segmentation (11 subjects) or automatic segmentation (17 subjects).

In order to generate parcellation labels, we used an automatic atlas-based method with an existing T2-weighted atlas of the fetal brain with verified parcellation labels \cite{gholipour2017normative}. To achieve accurate alignment of the atlas to the subject brain, we computed the registration transform as the composition of two transforms. First, the T2 atlas image was registered to the age-matched mean diffusivity atlas image using diffeomorphic registration \cite{avants2008symmetric}. Then, the fetus DTI image was registered to its age-matched atlas DTI image via DTI-based deformable registration\cite{zhang2006deformable}. If we denote these registration transforms, respectively, with $\Phi_1$ and $\Phi_2$, the composite transform is computed as $\Phi_c= \Phi_2^{-1} \circ \Phi_1$. The composite transform is applied to align the parcellation map from the T2 atlas to the individual fetus's brain in the dMRI space. The resulting parcellation maps were individually inspected, edited as needed, and verified by an expert. The same automatic procedure was applied to compute the parcellation label maps for both the training and test images. The names of the regions and the abbreviations used in this paper can be found in Table \ref{table:parcel_names} in the Supplementary Materials.

\subsection{Deep learning-based segmentation method}
\label{sec:segmentation_method}

\subsubsection{Network design}
\label{sec:MTL_method_design}

\begin{figure}[!ht]
\centering
\includegraphics[width=\linewidth]{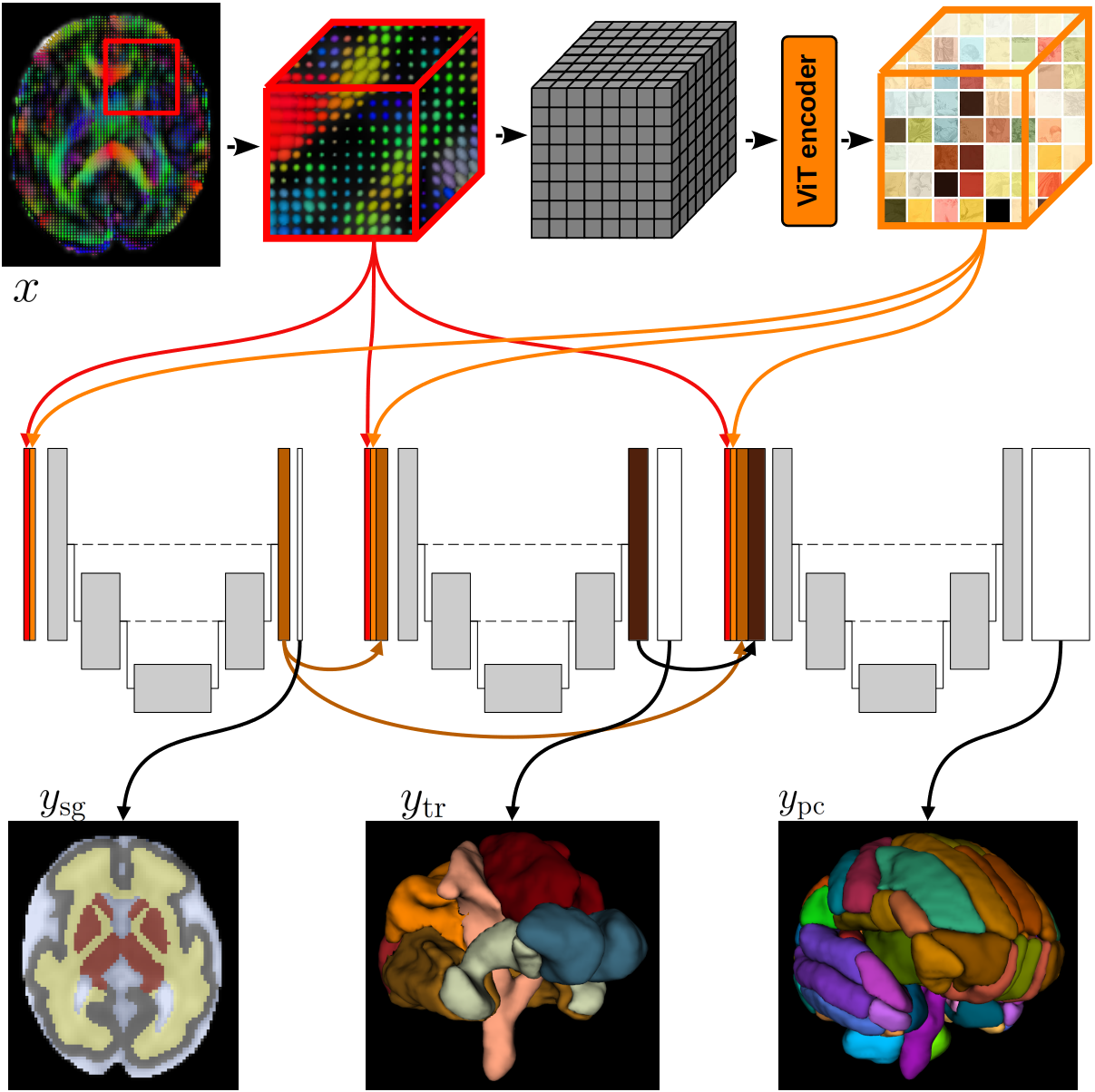}
\caption{Schematic representation of the multi-task deep learning framework. Given the DTI map as input ($x$), the model computes tissue segmentation ($y_{\text{sg}}$), white matter tract segmentations ($y_{\text{tr}}$), and parcellation ($y_{\text{pc}}$).}
\label{fig:method_outline}
\end{figure}

Our proposed automatic segmentation method is shown in Figure \ref{fig:method_outline}. We decided to use the DTI map computed from the dMRI data volume as the method input. This choice has the advantage of making the model generalizable to different dMRI acquisition schemes (i.e., gradient tables) that may vary greatly between scans. The diffusion tensor image can be estimated with typical fetal dMRI scans and does not require tailored multi-shell scans. Moreover, the DTI map contains the information about the primary fiber orientation, which would be lacking in scalar maps such as fractional anisotropy.

Our method consists of a cascade of networks that estimate the tissue segmentation, white matter tract segmentation, and parcellation, in that order. First, the DTI input is passed to a patch-wise attention module to facilitate the learning of long-range correlations early in the pipeline. The features learned by this module ($\text{F}_{\text{att}}$) are passed to a set of three fully convolutional networks (FCNs) that separately address the three segmentation tasks. There is no weight sharing among the FCNs. In addition to $\text{F}_{\text{att}}$, the DTI image is also included as the input to each FCN. Moreover, the penultimate FCN feature maps (just before the output layer) computed by each FCN are forwarded as additional input to the following FCN(s). A similar strategy has been adopted by at least one prior work in MTL. For semantic instance segmentation, Dai et al. \cite{dai2016instance} proposed a cascade of three networks for differentiating the instances, computing the segmentation masks, and object classification. In that work, the three sub-networks shared the same pool of convolutional features, and only the output of earlier tasks was passed to the following tasks. In our proposed method, on the other hand, there is no shared feature pool. Instead, the feature maps computed by earlier tasks are forwarded to the more downstream task(s). The only features that are shared among the tasks are those computed by the attention module. Moreover, each task computes new feature maps from the source DTI input. In summary, denoting the DTI map with $x$, the input to the tissue segmentation, WM tract segmentation, and parcellation FCNs are, respectively $[x; \text{F}_{\text{att}}]$, $[x; \text{F}_{\text{att}}; \text{F}_{\text{sg}}]$, and $[x; \text{F}_{\text{att}}; \text{F}_{\text{sg}}; \text{F}_{\text{tr}}]$, where $\text{F}_{\text{sg}}$ and $\text{F}_{\text{tr}}$ denote the penultimate feature maps from the tissue segmentation and WM tract segmentation FCNs.

The method works on cubes of size $64^3$ voxels. The whole cube is used as an input to the FCNs. For the attention module, the cube is partitioned into $8^3$ patches, each of size $8^3$ voxels. The attention module is identical to that in our prior work \cite{karimi2021convolution}. We used a succession of 4 vision transformer encoders with an embedding dimension of 512 and 4 attention heads. We refer to \cite{karimi2021convolution, dosovitskiy2020image} for architectural details. The output of the attention module is a set of six feature maps of the same spatial dimension as the input, i.e., $64^3$ voxels. These feature maps are passed to the three FCNs. For the FCNs, we used a standard U-Net design following the details of the nnU-Net \cite{isensee2021nnu}. We empirically set the number of feature maps in the first stage of each of the three FCNs to be 32. For other options, we generally followed the default settings of the nnU-Net framework \cite{isensee2021nnu}.

For tissue segmentation ($y_{\text{sg}}$) and parcellation ($y_{\text{pc}}$), the classes are mutually exclusive. In other words, each voxel belongs to one and only one class (considering the ``background'' as one of the labels). Therefore, a voxel-wise softmax function is applied to the output layer of the FCNs for these two tasks to compute the class probabilities. For tract segmentation ($y_{\text{tr}}$), on the other hand, the classes are not mutually exclusive because two or more tracts may cross the same voxel. Therefore, a sigmoid function is applied to the output layer of the tract segmentation FCN to compute an independent segmentation probability map for each tract.

\subsubsection{Multi-task learning approach and loss function}
\label{sec:MTL_approach}

Important considerations in the design and optimization of our method include the ordering, prioritization, and weighting of the tasks. These are persistent issues that have been extensively studied in MTL research. For most of these considerations, there are disagreements among prior studies and the optimal solution may depend on the specifics of the tasks involved and the performance targets. For example, in terms of task prioritization and loss weighting, some approaches such as curriculum learning \cite{bengio2009curriculum} and uncertainty-based weighting \cite{kendall2018multi} prioritize easier tasks, while other approaches, such as Dynamic Task Prioritization \cite{guo2018dynamic}, argue that more difficult tasks should be prioritized. In this work, we ordered the tasks in the cascade from easy to hard such that features learned from easier tasks were forwarded to the more difficult tasks. 

In order to determine which tasks were more difficult, we performed preliminary experiments where we trained separate models to address each of the three tasks. These experiments showed that, based on segmentation performance metrics such as the Dice Similarity Coefficient (DSC), the ordering of the tasks from easier to harder was: tissue segmentation, white matter tract segmentation, and parcellation. Our approach (Figure \ref{fig:method_outline}) allocates more of the model's capacity, in terms of the number of weights and computations, to the harder tasks. Moreover, the harder tasks in our application also have less reliable training labels. Specifically, in the case of cortical parcellation, there is a lack of consensus on cortical boundaries, especially in early fetal brain development. The smooth surface of the fetal brain during these stages (and therefore lack of gyrification) makes it difficult to set accurate boundaries, as the necessary anatomical landmarks are not fully developed. Hence, the features learned for the harder tasks are likely to be less useful for the easier tasks (e.g., tissue segmentation) that enjoy more accurate labels. Features learned by the earlier/easier tasks, on the other hand, are anticipated to be useful for the later/harder tasks. Specifically, tissue segmentation is highly informative for both WM tract segmentation and parcellation because the labels for these two later tasks are inherently dependent on the tissue type.

In order to balance the optimization loss for the three tasks, we learn task weightings based on homoscedastic (task-dependent) uncertainty as proposed in \cite{kendall2018multi}. This type of prediction uncertainty only depends on the task and does not vary as a function of the data instance. Rather than being a model output, it is encoded as a separate set of free parameters (one for each of the prediction labels). As shown below, these weights are included in the optimization loss function. Heteroscedastic (data-dependent) uncertainty may also be learned, for example via voxel-wise loss attenuation \cite{kendall2017}. However, this was not explored in our work because it would prohibitively increase the memory requirements due to the large number of prediction classes involved. Hence, the loss function used to optimize our model is as follows.

\begin{equation} \label{eq:loss}
\begin{split}
\mathcal{L}( x, y_{\text{sg}}, & y_{\text{tr}}, m_{\text{tr}}, y_{\text{pc}} | \theta, w)= \\ 
& - \mathlarger{\sum}_{i=1:l_{\text{sg}}} \exp(-w_{\text{sg}}^i) \cdot \text{DSC}( y_{\text{sg}}^i, \hat{y}_{\text{sg}}^i ) \\ 
 & - \mathlarger{\sum}_{i=1:l_{\text{tr}}} \exp(-w_{\text{tr}}^i) \cdot m_{\text{tr}}^i \cdot \text{DSC}( y_{\text{tr}}^i, \hat{y}_{\text{tr}}^i )  \\ 
 & - \mathlarger{\sum}_{i=1:l_{\text{pc}}} \exp(-w_{\text{pc}}^i) \cdot \text{DSC}( y_{\text{pc}}^i, \hat{y}_{\text{pc}}^i ) \\
 & + \mathlarger{\sum}_{i=1:l_{\text{sg}}} w_{\text{sg}}^i + \mathlarger{\sum}_{i=1:l_{\text{tr}}} m_{\text{tr}}^i \cdot w_{\text{tr}}^i + \mathlarger{\sum}_{i=1:l_{\text{pc}}} w_{\text{pc}}^i
\end{split}
\end{equation}

In this equation, the subscripts ``sg'', ``tr'', and ``pc'' respectively denote the variables for tissue segmentation, white matter tract segmentation, and parcellation. Symbols $y$ and $\hat{y}$ denote the target label and model prediction, respectively, $l_*$ is the number of classes for each task, and $\exp(-w_*)$ is the task weight in the loss function. In order to ignore the tracts that are missing in the training data, a zero-one array is introduced, which we have denoted with $m_{\text{tr}}$. Parameters to be optimized include the network weights ($\theta$), which includes the attention model and the three FCNs, and the task weights ($w$).

\subsubsection{Training and evaluation details}
\label{sec:training_and_validation}

During training, patches from random locations in the training images are used to optimize the model. Given the large model size, we used a batch size of 1. The model was optimized using stochastic gradient descent with an initial learning rate of 0.0001. We reduced the learning rate by a factor of 0.90 when the validation loss did not improve after 3 consecutive epochs. On a test image, we applied the model in a sliding window fashion with 25\% window overlap (i.e., 16 voxels) in each direction. The complete training code and the trained model weights are available at \url{https://github.com/engineeringmath/fetal_dmri_delineation}. A link to a Docker image of our final model is also available at the same repository.

We refer to the proposed multi-task method (Figure \ref{fig:method_outline}) simply as MTL. For comparison, we trained individual models separately for each task. In this approach, the model would look the same as what Figure \ref{fig:method_outline} shows for the first task (i.e., tissue segmentation task), but repeated for each task separately. We refer to this approach as single-task learning (STL). We also compare with nnU-Net \cite{isensee2021nnu} and a purely attention-based model \cite{karimi2021convolution} that we refer to as ATT. For quantitative evaluations, we use DSC, 95-percentile of the Hausdorff Distance (HD95), and Average Surface Distance (ASD). Given the challenges in accurately defining boundaries for cortical parcellation, especially due to the absence of gyrification in early fetal development, the parcellation labels were inherently less reliable. As a result, parcellation results were additionally reviewed by a human expert to accommodate this variability.

\section{Results and Discussion}
\label{sec:results_and_discussion}

\begin{figure}[!ht]
\centering
\includegraphics[width=\linewidth]{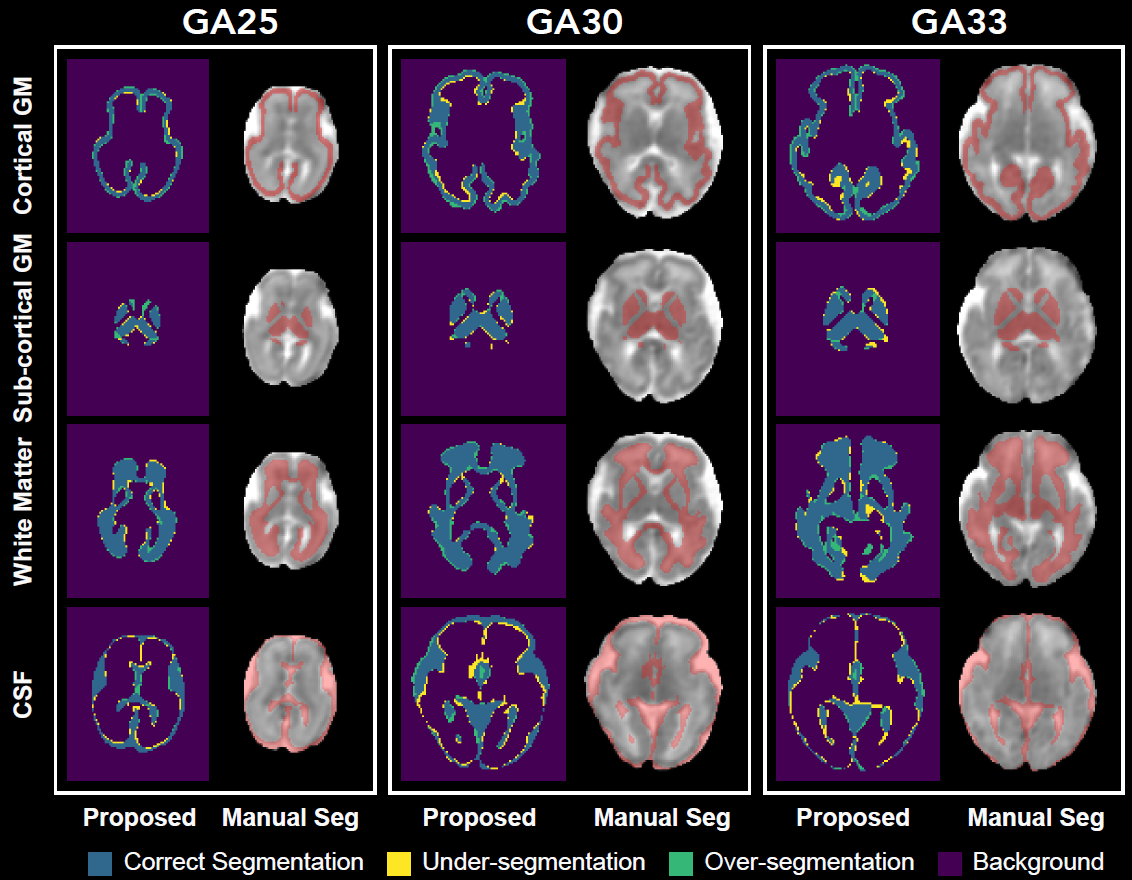}
\caption{Example tissue segmentation maps predicted by the proposed method for three test fetuses at 25, 30, and 33 gestational weeks.}
\label{fig:sg_example_results}
\end{figure}

\begin{figure}[!ht]
\centering
\includegraphics[width=\linewidth]{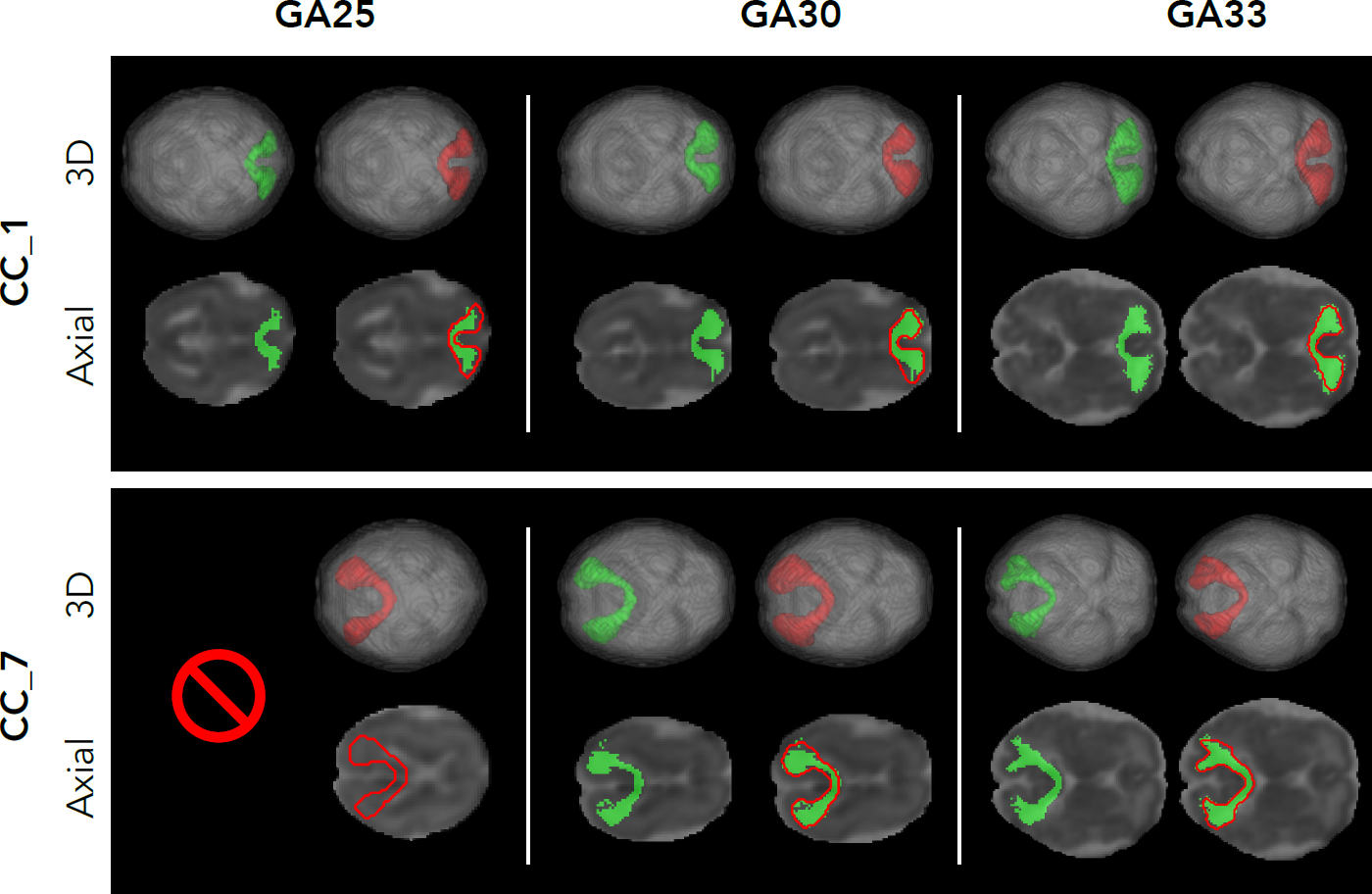}
\caption{Example commissural tract segmentation masks predicted by the proposed method for three test fetuses at 25, 30, and 33 weeks of gestation. Green shows the reference tract, and red shows the segmentation computed by the proposed method. In this example, for the fetus at 25 gestational weeks, the tractography-based method used to generate the ``ground truth'' failed.}
\label{fig:sg_example_results_commissural}
\end{figure}

\begin{figure}[!ht]
\centering
\includegraphics[width=\linewidth]{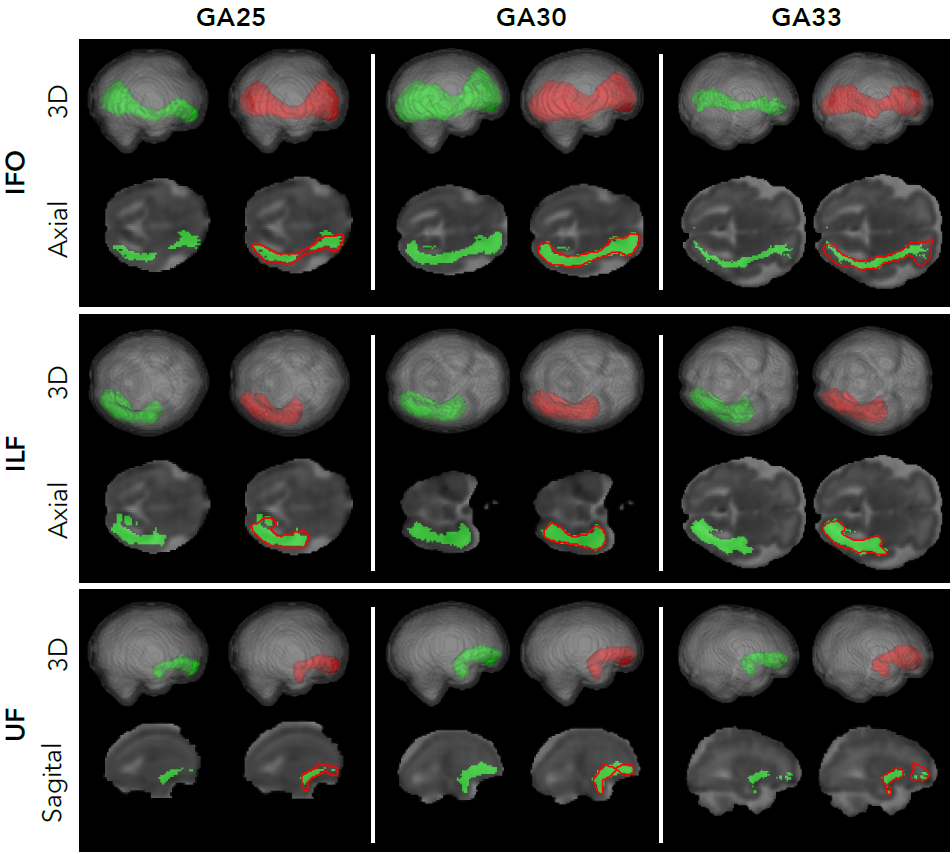}
\caption{Example association tract segmentation masks predicted by the proposed method for three test fetuses at 25, 30, and 33 weeks of gestation. Green shows the reference tract, and red shows the segmentation computed by the proposed method.}
\label{fig:sg_example_results_association}
\end{figure}

\begin{figure}[!ht]
\centering
\includegraphics[width=\linewidth]{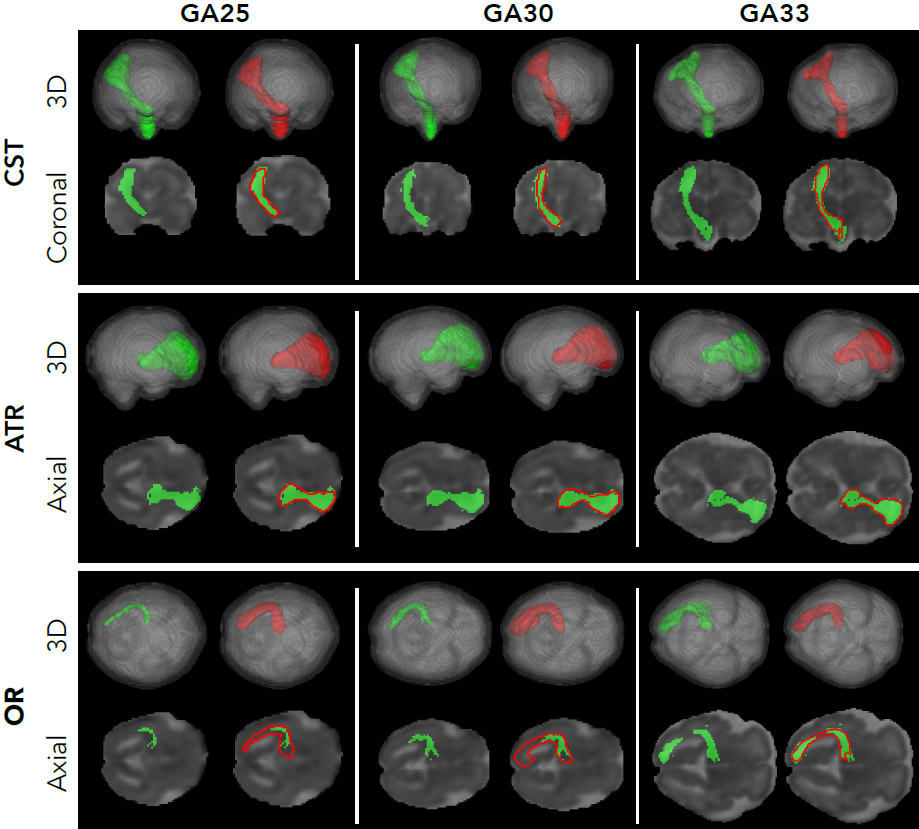}
\caption{Example projection tract segmentation masks predicted by the proposed method for three test fetuses at 25, 30, and 33 weeks of gestation. Green shows the reference tract, and red shows the segmentation computed by the proposed method.}
\label{fig:sg_example_results_projection}
\end{figure}

\begin{figure}[!ht]
\centering
\includegraphics[width=\linewidth]{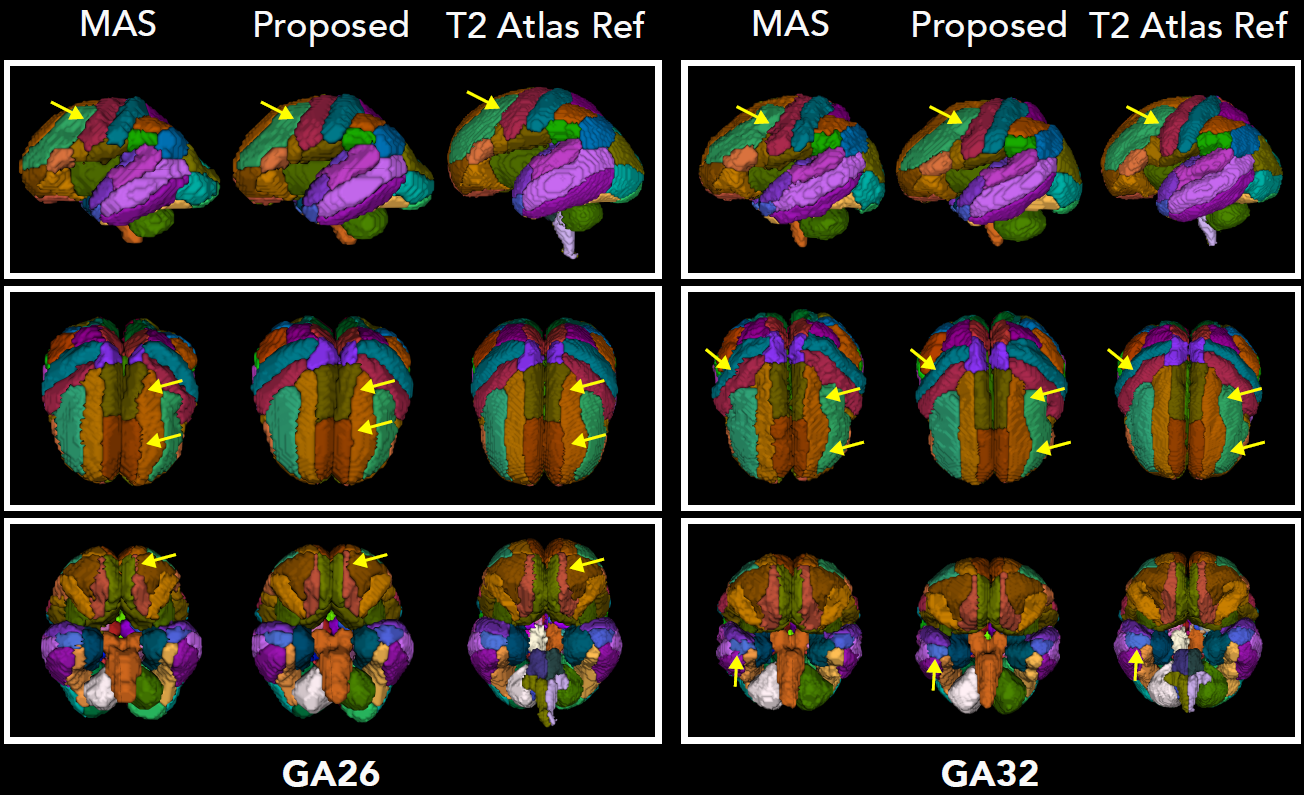}
\caption{Example brain parcellations computed by the proposed method on two test fetuses at 26 and 32 gestational weeks. The parcellation computed with the multi-atlas segmentation (MAS) method, used to generate labels on the training images, is shown. To allow for a visual comparison, we have also included the parcellation map of an age-equivalent T2 atlas image. Arrows highlight specific locations where our method's predictions outperform the MAS-quality labels that were used for model training.}
\label{fig:pc_example_results}
\end{figure}

Figures \ref{fig:sg_example_results} to \ref{fig:pc_example_results} show example tissue segmentation, WM tract segmentation, and parcellations computed by our model on independent test images. For WM tract segmentation, we have shown examples for commissural, association, and projection tracts separately in Figures \ref{fig:sg_example_results_commissural}, \ref{fig:sg_example_results_association}, and \ref{fig:sg_example_results_projection}. Despite the small size of our training dataset, exceptionally low image contrast, and rapidly developing brain structures with gestational age, our method has learned to compute accurate predictions for all three tasks across the gestational age.

Table \ref{table:summary_results} presents a summary of the segmentation performance metrics for the proposed method and the three compared techniques. The values in this table represent the average and standard deviation across all labels for each of the three tasks. A comparison of MTL and STL demonstrates the effectiveness of the proposed multi-task learning approach in improving the segmentation accuracy. The MTL approach has resulted in statistically significant improvements in the segmentation performance for all three tasks. Even for the tissue segmentation task, which is the first task in our framework and has the most reliable labels, there was a statistically significant improvement in DSC and HD95 compared with STL. This indicates that the information supplied by the tract segmentation and parcellation labels was effective in improving the training of the tissue segmentation FCN. Although the features learned by the tract segmentation and parcellation FCNs are not used by the tissue segmentation FCN, the model is trained as a whole. In other words, as the loss function in Equation \ref{eq:loss} shows, all network weights ($\theta$) are optimized with respect to the three tasks. The optimization gradient from the WM tract segmentation and parcellation tasks have benefited the training of the tissue segmentation FCN as well. For WM tract segmentation and parcellation tasks, the proposed MTL method has achieved significantly better performance metrics than STL and the two alternative methods. Compared with STL, the MTL approach has improved the DSC by 2.2\% on both tasks, which is statistically significant.

\begin{table*}[!htb]
\centering
\caption{Comparison of the proposed multi-task learning method (MTL) with single-task learning approach (STL) and two alternative deep learning-based segmentation techniques. The values shown are the average $\pm$ standard deviation across all class labels in each task. Bold results are statistically better than non-bold results in each row, where we have used paired t-tests with $p=0.01$.}
\label{table:summary_results}
\begin{tabular}{ | L{30mm} | L{10mm} | C{22mm} C{22mm} C{20mm} C{20mm} | }
\hline
Task & Metric & MTL & STL & nnU-Net & ATT \\ \hline
\multirow{3}{*}{Tissue segmentation} & DSC & $\mathbf{0.865 \pm 0.038}$ & $0.857 \pm 0.051$ & $0.843 \pm 0.049$ & $0.845 \pm 0.061$ \\
& HD95 & $\mathbf{1.252 \pm 0.271}$ & $1.366 \pm 0.305$ & $1.372 \pm 0.313$ & $1.358 \pm 0.307$ \\
& ASD & $\mathbf{0.344 \pm 0.068}$ & $\mathbf{0.344 \pm 0.073}$ & $0.357 \pm 0.071$ & $0.350 \pm 0.069$ \\   \hline
\multirow{3}{*}{Tract segmentation} & DSC & $\mathbf{0.825 \pm 0.051}$ & $0.803 \pm 0.061$ & $0.794 \pm 0.064$ & $0.801 \pm 0.072$ \\
& HD95 & $\mathbf{1.001 \pm 0.218}$ & $1.087 \pm 0.234$ & $1.116 \pm 0.241$ & $1.096 \pm 0.238$ \\
& ASD & $\mathbf{0.338 \pm 0.070}$ & $0.349 \pm 0.082$ & $0.350 \pm 0.81$ & $0.345 \pm 0.083$ \\   \hline
\multirow{3}{*}{Parcellation} & DSC & $\mathbf{0.819 \pm 0.070}$ & $0.797 \pm 0.074$ & $0.792 \pm 0.075$ & $0.766 \pm 0.106$ \\
& HD95 & $\mathbf{1.164 \pm 0.282}$ & $1.213 \pm 0.298$ & $1.216 \pm 0.294$ & $1.247 \pm 0.319$ \\
& ASD & $\mathbf{0.360 \pm 0.123}$ & $0.383 \pm 0.137$ & $0.391 \pm 0.132$ & $0.402 \pm 0.152$ \\   \hline
\end{tabular}
\end{table*}

\begin{figure*}[!ht]
\centering
\includegraphics[width=\textwidth]{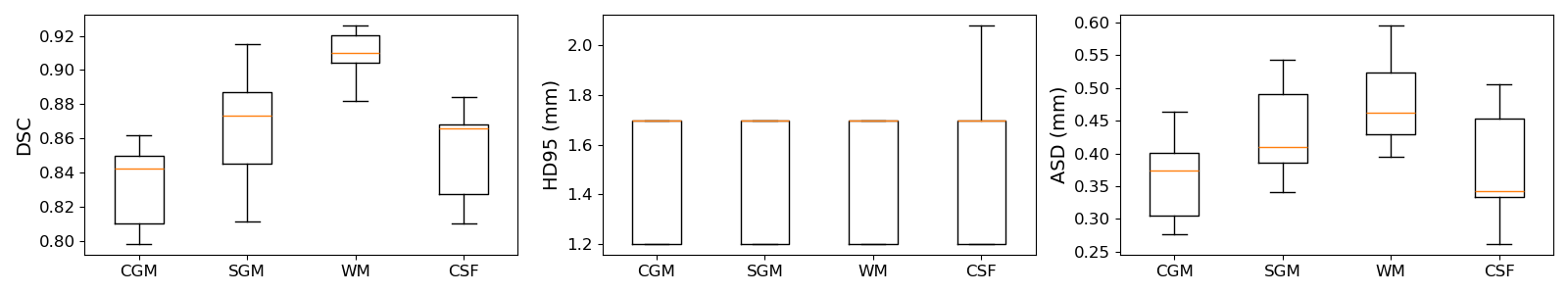}
\caption{Detailed segmentation performance metrics for the proposed MTL method in the tissue segmentation task.}
\label{fig:seg_results_detailed}
\end{figure*}

Figures \ref{fig:seg_results_detailed} shows the segmentation accuracy metrics for our method (MTL) for different class labels in the tissue segmentation task. Our method has achieved higher DSC for WM and SGM than for CGM and CSF. However, our method also shows higher ASD for WM and SGM. This suggests that the method has overall less accuracy in delineating the boundaries between WM and its neighboring tissues than delineating the boundaries between CSF and its neighboring tissues. The higher DSC for WM compared with CSF is due to the overall much larger volume of WM.

\begin{figure*}[!ht]
\centering
\includegraphics[width=\textwidth]{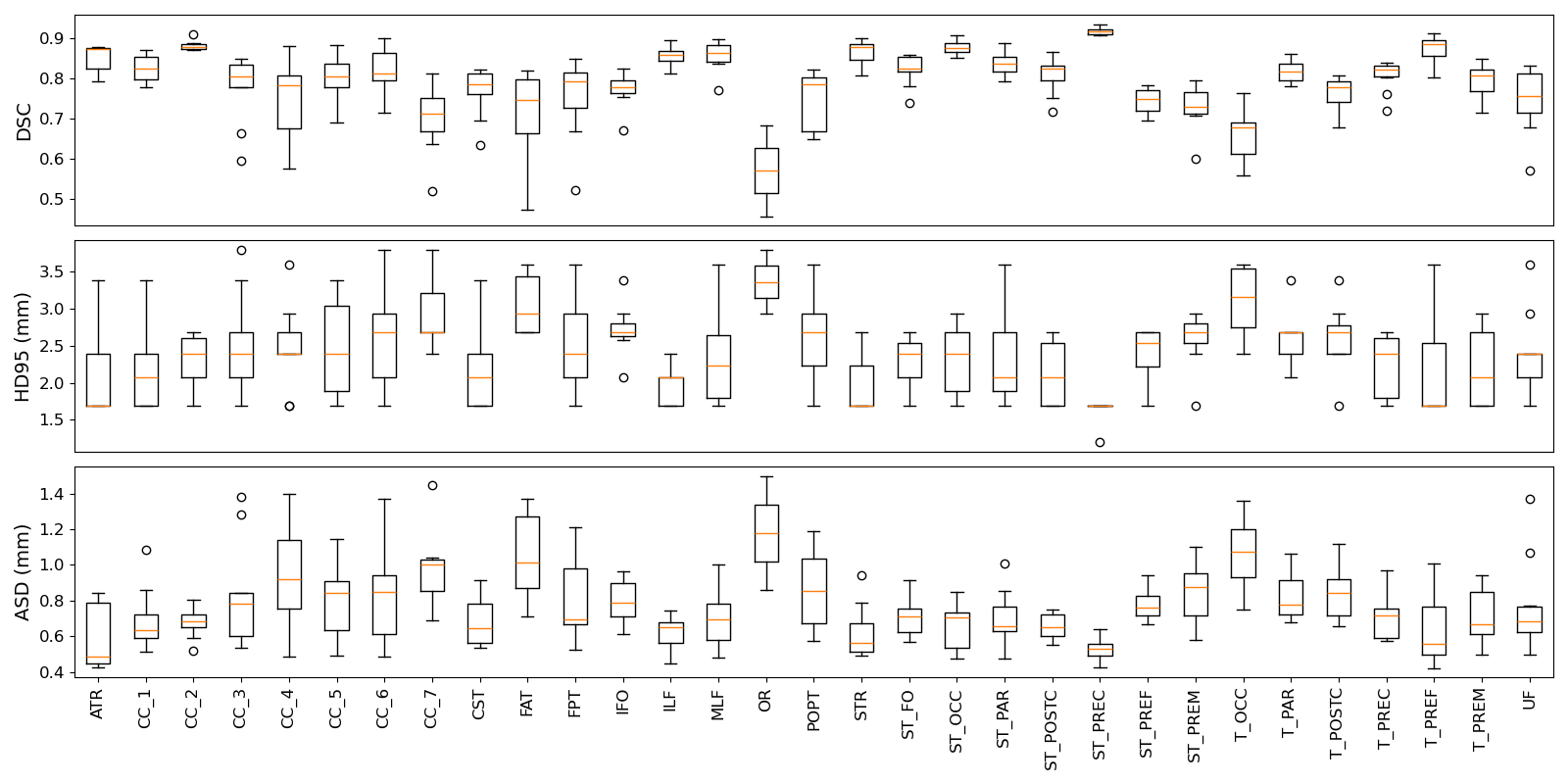}
\caption{Segmentation performance metrics for the proposed MTL method for each of the white matter tracts. Description of abbreviated tract names can be found in Table \ref{table:tract_names}.}
\label{fig:WM_tract_results_detailed}
\end{figure*}

Because our work is the first to compute fetal tissue segmentation in dMRI, we cannot directly compare our results with prior works. Nonetheless, it can be instructive to compare our results with the state of the art in (1) fetal tissue segmentation in anatomical MRI and (2) adult brain segmentation in dMRI. Despite fetal dMRI having lower tissue contrast than fetal T2-weighted MRI, our results are on par with the state of the art in fetal tissue segmentation in T2-weighted images \cite{payette2023fetal, fidon2021distributionally, dou2020deep, dumast2021segmentation}. Methods that have been designed specifically to segment the cortical plate have reported mean DSC values between 0.81 and 0.91 \cite{karimi2021transfer, fidon2021distributionally, dou2020deep, dumast2021segmentation, karimi2022improving}. Recent studies have shown mean DSC values of around 0.86, 0.72, and 0.78 for segmenting WM, CGM, and SGM, respectively on fetal structural MRI \cite{payette2023fetal}. For extra-axial CSF and the ventricles, the highest mean DSC has been approximately 0.85 and 0.90, respectively \cite{payette2023fetal}. Overall, our new method yields results that are comparable with or better than those results. For segmenting GM, WM, and CSF in adult brain dMRI, prior works have reported DSC values, respectively, in the range [0.68,0.86], [0.80,0.87], and [0.60,0.83] \cite{yap2015brain, ciritsis2018automated, zhang2015deep}, with one recent study reporting DSC values exceeding 0.95 on high-quality HCP data \cite{zhang2021deepseg}. Therefore, our results are also comparable with or better than most published works on adult brains.

Figure \ref{fig:WM_tract_results_detailed} shows the segmentation performance of the proposed MTL method for individual white matter tracts. The average DSC varies substantially for different tracts. The lowest DSC is observed for optic radiation (OR, DSC=0.57) and thalamo-occipital radiation (T\_OCC, DSC= 0.68), whereas the highest DSC was for prefrontal-striatal (ST\_PREC, DSC=0.92) and several tracts that had a DSC of approximately 0.88 including genu of the corpus callosum (CC\_2), superior thalamic radiation (STR), occipito-striatal (ST\_OCC), and thalamo-prefrontal radiation (T\_PREF). The lower accuracy for some of the tracts, such as OR, was mainly due to the high variability and low accuracy in the training labels. Specifically, due to the limited accuracy of fetal brain tractography, tracts such as OR were often only partially reconstructed.

\begin{figure*}[!ht]
\centering
\includegraphics[width=\textwidth]{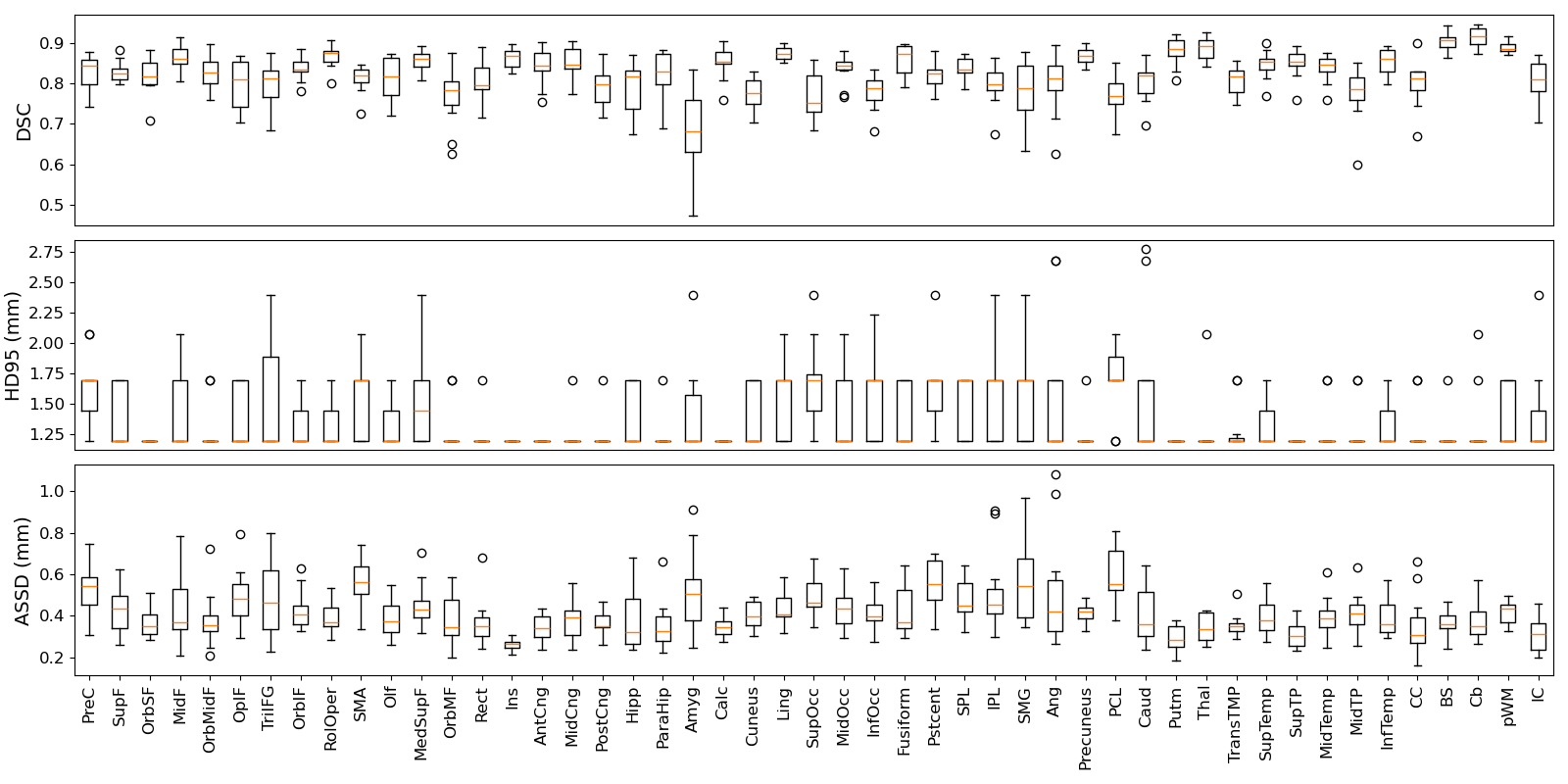}
\caption{Segmentation performance metrics for the proposed MTL method for the gray matter parcellation task. The full names of the regions shown on the horizontal axis can be found in Table \ref{table:parcel_names}.}
\label{fig:parcellation_results_detailed}
\end{figure*}

A comparison of our results with recent studies on adult brains shows that the performance of our method is highly competitive with those studies \cite{wasserthal2018tractseg, li2020neuro4neuro, reisert2018hamlet}. For instance, TractSeg \cite{wasserthal2018tractseg}, which is considered to be the sate of the art method, has reported a mean DSC of 0.84 for high-quality HCP data and 0.82 for clinical data. The mean tract DSC for TractSeg ranged from 0.63 to 0.90, with the highest DSC observed for the genu of the corpus callosum (CC\_2), similar to our results. TractSeg is able to segment several smaller tracts, such as the fornix and anterior commissure, which are not included in our data because we were unable to reconstruct them in our fetal whole-brain tractograms. Nonetheless, this comparison shows that our method is comparable with the state-of-the-art methods for adult brains. 

Figure \ref{fig:parcellation_results_detailed} shows the region-specific performance metrics for our method in the parcellation task. Overall, the proposed method shows consistent average performance for all parcellation regions. The mean DSC for most regions is close to or higher than 0.80, and the mean ASSD for most regions is close to or lower than 0.50mm. The amygdala (Amyg) had the lowest DSC (0.68), while all other regions had a DSC above 0.75. Only three regions, namely the Supplementary Motor Area (SMA), Postcentral Gyrus (Pstcent), and Paracentral Lobule (PCL), had a mean ASSD higher than 0.50mm. The cortical boundaries for these regions, as well as all other cortical areas, are exceptionally challenging to define in the fetus, even when done manually. As a result, it is important to interpret the quantitative performance metrics for the parcellation task with caution.

We are unaware of any prior works on fetal brain parcellation in dMRI. For parcellation of cortical regions in T2-weighted fetal MRI, one study has reported a mean DSC of 0.899 and median surface distance of 2.47mm \cite{you2024automatic}. For adult brain parcellation in dMRI, on the other hand, a recent deep learning method achieved a mean DSC of 0.76 and a mean test-retest DSC of 0.86. A direct comparison of our results with those studies is nontrivial. In fact, a fair comparison across such disparate settings may be impossible. Tissue contract in dMRI is very different from T2-weighted images, and the fetal brain images are characterized by low signal quality and rapid brain development, which makes them far more challenging than adult brains. Moreover, as mentioned above, the parcellation labels are not as accurate as the tissue segmentation and WM tract segmentation labels. This limitation probably applies to all prior works on brain parcellation to different degrees. Therefore, quantitative assessments that compare the method predictions with the ``ground truth'' provide only a limited view of performance. In this work, in addition to the quantitative metrics reported above, an expert visually assessed the quality of each parcellation prediction by verifying its boundaries against known anatomical landmarks. For example, for the precentral gyrus, landmarks included the superior and inferior precentral sulci anteriorly, the central sulcus posteriorly, the medial longitudinal fissure medially, and the lateral sulcus laterally. As shown in Figure \ref{fig:pc_example_results}, the parcellations computed by the proposed method were, in some cases,  free from the local errors that existed in the training labels. This is a common behavior of deep learning models trained on datasets with noisy labels \cite{rolnick2017}.

Task weights (Equation \ref{eq:loss}) for our trained model were $w_{\text{sg}}=-9.43 \pm 0.01$, $w_{\text{tr}}= -8.91 \pm 0.23$, and $w_{\text{pc}}=-9.35 \pm 0.05$. These values indicate that the method assigned higher uncertainty values to the tract segmentation task compared to the other tasks. This can be because, for the tract segmentation, each label (i.e., each tract) is optimized separately, while for the other two tasks, there is a strict condition that labels are mutually exclusive. In other words, while neighboring labels influence each other in tissue segmentation and parcellation, this dependency does not exist for tract segmentation resulting in higher uncertainty in the predictions.

Despite the uniqueness of this study and the highly promising results, we acknowledge certain limitations that future works may address. Above, we pointed out the limited accuracy of the parcellation labels. In addition, the white matter tract segmentation labels were not complete because our annotation approach (described in Section \ref{sec:annotation}) was not able to consistently reconstruct all of them. Some of these tracts such as arcuate fasciculus, may not be fully developed in the fetal brain, while others such as the cingulum certainly exist but are challenging to automatically annotate. Improving the annotation procedures is likely to lead to further segmentation performance improvements compared with the results reported in this paper.

\section{Conclusions}

This work represents the first attempt to develop an automatic method for delineating fetal brain tissue and structures directly in the dMRI space. Given the small dataset size and the difficulty of obtaining annotations, we developed a multi-task learning approach to optimally leverage the training labels for three separate but related tasks. Our evaluations show that the proposed method can adequately address all three tasks in a unified computational framework and achieve performance levels that are comparable with the results reported in the most recent studies on adult brains and fetal structural MRI. Therefore, the new method can be used to automate and drastically facilitate the processing of fetal dMRI data. Due to the enormous potential of dMRI for studying early brain development, the new method can significantly enhance our ability to chart normal brain development in utero and to assess the impact of genetic and environmental factors that can disrupt this process.

\section*{Acknowledgements}

This research was supported in part by the National Institute of Neurological Disorders and Stroke under award number R01 NS128281; the Eunice Kennedy Shriver National Institute of Child Health and Human Development under award number R01 HD110772; and by NIH grants R01 LM013608, R01 EB019483, and R01 NS124212. The content of this publication is solely the responsibility of the authors and does not necessarily represent the official views of the NIH.

\bibliographystyle{ieeetr}
\bibliography{davoodreferences}

\begin{thebibliography}{10}

\bibitem{kostovic2006development}
I.~Kostovi{\'c} and N.~Jovanov-Milo{\v{s}}evi{\'c}, ``The development of
  cerebral connections during the first 20--45 weeks’ gestation,'' {\em
  Seminars in Fetal and Neonatal Medicine}, vol.~11, no.~6, pp.~415--422, 2006.

\bibitem{donofrio2011impact}
M.~T. Donofrio, C.~Limperopoulos, {\em et~al.}, ``Impact of congenital heart
  disease on fetal brain development and injury,'' {\em Current opinion in
  pediatrics}, vol.~23, no.~5, pp.~502--511, 2011.

\bibitem{lynch2009epidemiology}
J.~K. Lynch, ``Epidemiology and classification of perinatal stroke,'' {\em
  Seminars in Fetal and Neonatal Medicine}, vol.~14, no.~5, pp.~245--249, 2009.

\bibitem{egana2013differences}
G.~Ega{\~n}a-Ugrinovic {\em et~al.}, ``Differences in cortical development
  assessed by fetal mri in late-onset intrauterine growth restriction,'' {\em
  American journal of obstetrics and gynecology}, vol.~209, no.~2, pp.~126--e1,
  2013.

\bibitem{gholipour2014fetal}
A.~Gholipour {\em et~al.}, ``Fetal mri: a technical update with educational
  aspirations,'' {\em Concepts in Magnetic Resonance Part A}, vol.~43, no.~6,
  pp.~237--266, 2014.

\bibitem{ouyang2019delineation}
M.~Ouyang {\em et~al.}, ``Delineation of early brain development from fetuses
  to infants with diffusion mri and beyond,'' {\em Neuroimage}, vol.~185,
  pp.~836--850, 2019.

\bibitem{huang2009anatomical}
H.~Huang {\em et~al.}, ``Anatomical characterization of human fetal brain
  development with diffusion tensor magnetic resonance imaging,'' {\em Journal
  of Neuroscience}, vol.~29, no.~13, pp.~4263--4273, 2009.

\bibitem{khan2019fetal}
S.~Khan {\em et~al.}, ``Fetal brain growth portrayed by a spatiotemporal
  diffusion tensor mri atlas computed from in utero images,'' {\em NeuroImage},
  vol.~185, pp.~593--608, 2019.

\bibitem{calixto2024detailed}
C.~Calixto {\em et~al.}, ``A detailed spatio-temporal atlas of the white matter
  tracts for the fetal brain,'' {\em bioRxiv}, 2024.

\bibitem{dubois2014early}
J.~Dubois {\em et~al.}, ``The early development of brain white matter: a review
  of imaging studies in fetuses, newborns and infants,'' {\em Neuroscience},
  vol.~276, pp.~48--71, 2014.

\bibitem{jakab2017utero}
A.~Jakab {\em et~al.}, ``In utero diffusion tensor imaging of the fetal brain:
  a reproducibility study,'' {\em NeuroImage: Clinical}, vol.~15, pp.~601--612,
  2017.

\bibitem{karimi2024diffusion}
D.~Karimi, ``Diffusion mri with machine learning,'' {\em arXiv preprint
  arXiv:2402.00019}, 2024.

\bibitem{payette2023fetal}
K.~Payette {\em et~al.}, ``Fetal brain tissue annotation and segmentation
  challenge results,'' {\em Medical image analysis}, vol.~88, p.~102833, 2023.

\bibitem{calixto2023detailed}
C.~Calixto {\em et~al.}, ``Detailed anatomic segmentations of a fetal brain
  diffusion tensor imaging atlas between 23 and 30 weeks of gestation,'' {\em
  Human Brain Mapping}, vol.~44, no.~4, pp.~1593--1602, 2023.

\bibitem{karimi2023learning}
D.~Karimi {\em et~al.}, ``Learning to segment fetal brain tissue from noisy
  annotations,'' {\em Medical Image Analysis}, vol.~85, p.~102731, 2023.

\bibitem{smith2012anatomically}
R.~E. Smith {\em et~al.}, ``Anatomically-constrained tractography: improved
  diffusion mri streamlines tractography through effective use of anatomical
  information,'' {\em Neuroimage}, vol.~62, no.~3, pp.~1924--1938, 2012.

\bibitem{st2018surface}
E.~St-Onge {\em et~al.}, ``Surface-enhanced tractography (set),'' {\em
  Neuroimage}, vol.~169, pp.~524--539, 2018.

\bibitem{poulin2019tractography}
P.~Poulin {\em et~al.}, ``Tractography and machine learning: Current state and
  open challenges,'' {\em Magnetic resonance imaging}, vol.~64, pp.~37--48,
  2019.

\bibitem{rheault2020common}
F.~Rheault {\em et~al.}, ``Common misconceptions, hidden biases and modern
  challenges of dmri tractography,'' {\em Journal of neural engineering},
  vol.~17, no.~1, p.~011001, 2020.

\bibitem{wassermann2016white}
D.~Wassermann {\em et~al.}, ``The white matter query language: a novel approach
  for describing human white matter anatomy,'' {\em Brain Structure and
  Function}, vol.~221, pp.~4705--4721, 2016.

\bibitem{makropoulos2018review}
A.~Makropoulos, S.~J. Counsell, and D.~Rueckert, ``A review on automatic fetal
  and neonatal brain mri segmentation,'' {\em NeuroImage}, vol.~170,
  pp.~231--248, 2018.

\bibitem{dou2020deep}
H.~Dou {\em et~al.}, ``A deep attentive convolutional neural network for
  automatic cortical plate segmentation in fetal mri,'' {\em IEEE transactions
  on medical imaging}, vol.~40, no.~4, pp.~1123--1133, 2020.

\bibitem{dumast2021segmentation}
P.~d. Dumast {\em et~al.}, ``Segmentation of the cortical plate in fetal brain
  mri with a topological loss,'' in {\em Uncertainty for Safe Utilization of
  Machine Learning in Medical Imaging, and Perinatal Imaging, Placental and
  Preterm Image Analysis}, pp.~200--209, Springer, 2021.

\bibitem{khalili2019automatic}
N.~Khalili {\em et~al.}, ``Automatic brain tissue segmentation in fetal mri
  using convolutional neural networks,'' {\em Magnetic resonance imaging},
  vol.~64, pp.~77--89, 2019.

\bibitem{payette2020efficient}
K.~Payette, R.~Kottke, and A.~Jakab, ``Efficient multi-class fetal brain
  segmentation in high resolution mri reconstructions with noisy labels,'' in
  {\em Medical Ultrasound, and Preterm, Perinatal and Paediatric Image
  Analysis}, pp.~295--304, Springer, 2020.

\bibitem{you2024automatic}
S.~You {\em et~al.}, ``Automatic cortical surface parcellation in the fetal
  brain using attention-gated spherical u-net,'' {\em Frontiers in
  Neuroscience}, vol.~18, p.~1410936, 2024.

\bibitem{payette2021automatic}
K.~Payette {\em et~al.}, ``An automatic multi-tissue human fetal brain
  segmentation benchmark using the fetal tissue annotation dataset,'' {\em
  Scientific Data}, vol.~8, no.~1, pp.~1--14, 2021.

\bibitem{fetit2020deep}
A.~E. Fetit {\em et~al.}, ``A deep learning approach to segmentation of the
  developing cortex in fetal brain mri with minimal manual labeling,'' in {\em
  Medical Imaging with Deep Learning}, pp.~241--261, PMLR, 2020.

\bibitem{caruana1997multitask}
R.~Caruana, ``Multitask learning,'' {\em Machine learning}, vol.~28,
  pp.~41--75, 1997.

\bibitem{crawshaw2020multi}
M.~Crawshaw, ``Multi-task learning with deep neural networks: A survey,'' {\em
  arXiv preprint arXiv:2009.09796}, 2020.

\bibitem{vandenhende2021multi}
S.~Vandenhende {\em et~al.}, ``Multi-task learning for dense prediction tasks:
  A survey,'' {\em IEEE transactions on pattern analysis and machine
  intelligence}, vol.~44, no.~7, pp.~3614--3633, 2021.

\bibitem{zhang2014facial}
Z.~Zhang, P.~Luo, C.~C. Loy, and X.~Tang, ``Facial landmark detection by deep
  multi-task learning,'' in {\em Computer Vision--ECCV 2014: 13th European
  Conference, Zurich, Switzerland, September 6-12, 2014, Proceedings, Part VI
  13}, pp.~94--108, Springer, 2014.

\bibitem{misra2016cross}
I.~Misra, A.~Shrivastava, A.~Gupta, and M.~Hebert, ``Cross-stitch networks for
  multi-task learning,'' in {\em Proceedings of the IEEE conference on computer
  vision and pattern recognition}, pp.~3994--4003, 2016.

\bibitem{dai2016instance}
J.~Dai, K.~He, and J.~Sun, ``Instance-aware semantic segmentation via
  multi-task network cascades,'' in {\em Proceedings of the IEEE conference on
  computer vision and pattern recognition}, pp.~3150--3158, 2016.

\bibitem{moeskops2016deep}
P.~Moeskops {\em et~al.}, ``Deep learning for multi-task medical image
  segmentation in multiple modalities,'' in {\em Medical Image Computing and
  Computer-Assisted Intervention--MICCAI 2016: 19th International Conference,
  Athens, Greece, October 17-21, 2016, Proceedings, Part II 19}, pp.~478--486,
  Springer, 2016.

\bibitem{karimi2022improving}
D.~Karimi and A.~Gholipour, ``Improving calibration and out-of-distribution
  detection in deep models for medical image segmentation,'' {\em IEEE
  transactions on artificial intelligence}, vol.~4, no.~2, pp.~383--397, 2022.

\bibitem{liu2019loss}
S.~Liu, Y.~Liang, and A.~Gitter, ``Loss-balanced task weighting to reduce
  negative transfer in multi-task learning,'' in {\em Proceedings of the AAAI
  conference on artificial intelligence}, vol.~33, pp.~9977--9978, 2019.

\bibitem{fifty2021efficiently}
C.~Fifty {\em et~al.}, ``Efficiently identifying task groupings for multi-task
  learning,'' {\em Advances in Neural Information Processing Systems}, vol.~34,
  pp.~27503--27516, 2021.

\bibitem{kendall2018multi}
A.~Kendall {\em et~al.}, ``Multi-task learning using uncertainty to weigh
  losses for scene geometry and semantics,'' in {\em Proceedings of the IEEE
  conference on computer vision and pattern recognition}, pp.~7482--7491, 2018.

\bibitem{liu2019end}
S.~Liu, E.~Johns, and A.~J. Davison, ``End-to-end multi-task learning with
  attention,'' in {\em Proceedings of the IEEE/CVF conference on computer
  vision and pattern recognition}, pp.~1871--1880, 2019.

\bibitem{jean2019adaptive}
S.~Jean, O.~Firat, and M.~Johnson, ``Adaptive scheduling for multi-task
  learning,'' {\em arXiv preprint arXiv:1909.06434}, 2019.

\bibitem{vandenhende2019branched}
S.~Vandenhende {\em et~al.}, ``Branched multi-task networks: deciding what
  layers to share,'' {\em arXiv preprint arXiv:1904.02920}, 2019.

\bibitem{duong2015low}
L.~Duong {\em et~al.}, ``Low resource dependency parsing: Cross-lingual
  parameter sharing in a neural network parser,'' in {\em Proceedings of the
  53rd annual meeting of the Association for Computational Linguistics and the
  7th international joint conference on natural language processing (volume 2:
  short papers)}, pp.~845--850, 2015.

\bibitem{chaudhry2018efficient}
A.~Chaudhry, M.~Ranzato, M.~Rohrbach, and M.~Elhoseiny, ``Efficient lifelong
  learning with a-gem,'' {\em arXiv preprint arXiv:1812.00420}, 2018.

\bibitem{bengio2009curriculum}
Y.~Bengio {\em et~al.}, ``Curriculum learning,'' in {\em Proceedings of the
  26th annual international conference on machine learning}, pp.~41--48, 2009.

\bibitem{guo2018dynamic}
M.~Guo {\em et~al.}, ``Dynamic task prioritization for multitask learning,'' in
  {\em Proceedings of the European conference on computer vision (ECCV)},
  pp.~270--287, 2018.

\bibitem{marami2017temporal}
B.~Marami {\em et~al.}, ``Temporal slice registration and robust
  diffusion-tensor reconstruction for improved fetal brain structural
  connectivity analysis,'' {\em NeuroImage}, vol.~156, pp.~475--488, 2017.

\bibitem{zhang2006deformable}
H.~Zhang {\em et~al.}, ``Deformable registration of diffusion tensor mr images
  with explicit orientation optimization,'' {\em Medical image analysis},
  vol.~10, no.~5, pp.~764--785, 2006.

\bibitem{akhondi2013simultaneous}
A.~Akhondi-Asl and S.~K. Warfield, ``Simultaneous truth and performance level
  estimation through fusion of probabilistic segmentations,'' {\em IEEE
  transactions on medical imaging}, vol.~32, no.~10, pp.~1840--1852, 2013.

\bibitem{tournier2010improved}
J.~D. Tournier {\em et~al.}, ``Improved probabilistic streamlines tractography
  by 2nd order integration over fibre orientation distributions,'' in {\em
  Proceedings of the international society for magnetic resonance in medicine},
  vol.~1670, Ismrm, 2010.

\bibitem{gholipour2017normative}
A.~Gholipour {\em et~al.}, ``A normative spatiotemporal mri atlas of the fetal
  brain for automatic segmentation and analysis of early brain growth,'' {\em
  Scientific reports}, vol.~7, no.~1, pp.~1--13, 2017.

\bibitem{avants2008symmetric}
B.~B. Avants {\em et~al.}, ``Symmetric diffeomorphic image registration with
  cross-correlation: evaluating automated labeling of elderly and
  neurodegenerative brain,'' {\em Medical image analysis}, vol.~12, no.~1,
  pp.~26--41, 2008.

\bibitem{karimi2021convolution}
D.~Karimi, S.~D. Vasylechko, and A.~Gholipour, ``Convolution-free medical image
  segmentation using transformers,'' in {\em Medical image computing and
  computer assisted intervention--MICCAI 2021: 24th international conference,
  Strasbourg, France, September 27--October 1, 2021, proceedings, part I 24},
  pp.~78--88, Springer, 2021.

\bibitem{dosovitskiy2020image}
A.~Dosovitskiy {\em et~al.}, ``An image is worth 16x16 words: Transformers for
  image recognition at scale,'' {\em arXiv preprint arXiv:2010.11929}, 2020.

\bibitem{isensee2021nnu}
F.~Isensee {\em et~al.}, ``nnu-net: a self-configuring method for deep
  learning-based biomedical image segmentation,'' {\em Nature methods},
  vol.~18, no.~2, pp.~203--211, 2021.

\bibitem{kendall2017}
A.~Kendall and Y.~Gal, ``What uncertainties do we need in bayesian deep
  learning for computer vision?,'' in {\em Advances in neural information
  processing systems}, pp.~5574--5584, 2017.

\bibitem{fidon2021distributionally}
L.~Fidon {\em et~al.}, ``Distributionally robust segmentation of abnormal fetal
  brain 3d mri,'' in {\em Uncertainty for Safe Utilization of Machine Learning
  in Medical Imaging, and Perinatal Imaging, Placental and Preterm Image
  Analysis}, pp.~263--273, Springer, 2021.

\bibitem{karimi2021transfer}
D.~Karimi {\em et~al.}, ``Transfer learning in medical image segmentation: New
  insights from analysis of the dynamics of model parameters and learned
  representations,'' {\em Artificial Intelligence in Medicine}, vol.~116,
  p.~102078, 2021.

\bibitem{yap2015brain}
P.-T. Yap, Y.~Zhang, and D.~Shen, ``Brain tissue segmentation based on
  diffusion mri using l0 sparse-group representation classification,'' in {\em
  Medical Image Computing and Computer-Assisted Intervention--MICCAI 2015: 18th
  International Conference, Munich, Germany, October 5-9, 2015, Proceedings,
  Part III 18}, pp.~132--139, Springer, 2015.

\bibitem{ciritsis2018automated}
A.~Ciritsis, A.~Boss, and C.~Rossi, ``Automated pixel-wise brain tissue
  segmentation of diffusion-weighted images via machine learning,'' {\em NMR in
  Biomedicine}, vol.~31, no.~7, p.~e3931, 2018.

\bibitem{zhang2015deep}
W.~Zhang {\em et~al.}, ``Deep convolutional neural networks for multi-modality
  isointense infant brain image segmentation,'' {\em NeuroImage}, vol.~108,
  pp.~214--224, 2015.

\bibitem{zhang2021deepseg}
F.~Zhang {\em et~al.}, ``Deep learning based segmentation of brain tissue from
  diffusion mri,'' {\em NeuroImage}, vol.~233, p.~117934, 2021.

\bibitem{wasserthal2018tractseg}
J.~Wasserthal {\em et~al.}, ``Tractseg-fast and accurate white matter tract
  segmentation,'' {\em NeuroImage}, vol.~183, pp.~239--253, 2018.

\bibitem{li2020neuro4neuro}
B.~Li {\em et~al.}, ``Neuro4neuro: A neural network approach for neural tract
  segmentation using large-scale population-based diffusion imaging,'' {\em
  NeuroImage}, vol.~218, p.~116993, 2020.

\bibitem{reisert2018hamlet}
M.~Reisert {\em et~al.}, ``Hamlet: hierarchical harmonic filters for learning
  tracts from diffusion mri,'' {\em arXiv preprint arXiv:1807.01068}, 2018.

\bibitem{rolnick2017}
D.~Rolnick, A.~Veit, S.~Belongie, and N.~Shavit, ``Deep learning is robust to
  massive label noise,'' {\em arXiv preprint arXiv:1705.10694}, 2017.

\end{thebibliography}

\clearpage

\begin{table*}[!htb]
\footnotesize
\centering
\caption{The names and abbreviations of the WM tracts considered in this work. Detailed description and visual presentation of these tracts can be found in \cite{calixto2024detailed}.}
\label{table:tract_names}
\begin{tabular}{llll|llll}
\thickhline
\multicolumn{3}{l}{Classification / Subdivision / Tract}   & Abbr. & \multicolumn{3}{l}{Classification / Subdivision / Tract} & Abbr. \\
\thickhline
\multicolumn{3}{l}{Projection Tracts}                             &              & \multicolumn{3}{l}{Association Tracts}                          &              \\
 &  & Corticospinal tract                                  & CST          &     &    & Frontal Aslant tract                          & FAT          \\
 & \multicolumn{2}{l}{Cortico-ponto-cerebellar}            &              &     &    & Inferior fronto-occipital fasciculus          & IFO          \\
 &  & Fronto-pontine tract                                 & FPT          &     &    & Inferior longitudinal fascicle                & ILF          \\
 &  & Parieto-occipital pontine tract                      & POPT         &     &    & Middle longitudinal fascicle                  & MLF          \\
 & \multicolumn{2}{l}{Cortico-striatal}                    &              &     &    & Uncinate fasciculus   & UF           \\
 &  & Fronto-orbital-striatal                              & ST\_FO       & \multicolumn{3}{l}{Commisural Tracts}                           &              \\
 &  & Occipito-striatal                                    & ST\_OCC      &     & \multicolumn{2}{l}{Corpus Callosum}                &              \\
 &  & Parieto-striatal                                     & ST\_PAR      &     &    & Rostrum                                       & CC\_1        \\
 &  & Postcentral-striatal                                 & ST\_POSTC    &     &    & Genu                                          & CC\_2        \\
 &  & Precentral-striatal                                  & ST\_PREC    &     &    & Rostral body                                  & CC\_3        \\
 &  & Prefrontal-striatal                                  & ST\_PREF     &     &    & Anterior midbody                              & CC\_4        \\
 &  & Premotor-striatal                                    & ST\_PREM     &     &    & Posterior midbody                             & CC\_5        \\
 & \multicolumn{2}{l}{Thalamic radiations}                 &              &     &    & Isthmus                                       & CC\_6        \\
 &  & Anterior thalamic radiation                          & ATR          &     &    & Splenium                                      & CC\_7        \\
 &  & Optic radiation                                      & OR           &                         &              \\
 &  & Superior thalamic radiation                          & STR          &     &    &                 &         \\
 &  & Thalamo-occipital radiation  & T\_OCC       &     &    &                   &           \\
 &  & Thalamo-parietal radiation                           & T\_PAR       &     &    &               &         \\
 &  & Thalamo-postcentral radiation                        & T\_POSTC     &     &    &                                               &              \\
 &  & Thalamo-precentral radiation & T\_PREC      &     &    &                                               &              \\
 &  & Thalamo-prefrontal radiation                         & T\_PREF      &     &    &                                               &              \\
 &  & Thalamo-premotor radiation                           & T\_PREM      &     &    &                                               &             \\
 \thickhline
\end{tabular}
\end{table*}

\begin{table*}[]
\centering
\footnotesize
\caption{The full names and abbreviations for the brain regions and structures considered in the parcellation task.}
\label{table:parcel_names}
\begin{tabular}{lll|lll}
\thickhline
\multicolumn{2}{l}{Zone / Segmentation Name} & Abbr.                   & \multicolumn{2}{l}{Zone / Segmentation Name}         & Abbr.                     \\
\thickhline
\multicolumn{2}{l}{Frontal Lobe}                &                                & \multicolumn{2}{l}{Temporal Lobe}            &                                  \\
& {Gyrus Rectus}                                  & {Rect}    &          & {Temporal Pole (Superior)}           & {SupTP}     \\
& {Medial Superior Frontal Gyrus}                 & {MedSupF} &          & Hippocampus                                               & Hipp                             \\
& {Middle Frontal Gyrus}                          & {MidF}    &          & {Inferior Temporal Gyrus}            & {InfTemp}   \\
& {Olfactory Cortex}                              & Olf                            &          & {Middle Temporal Gyrus}              & {MidTemp}   \\
& {Opercular Part of the Inferior Frontal Gyrus}  & OpIF                           &          & {Parahippocampal Gyrus}              & {ParaHip}   \\
& {Orbital Part of the Inferior Frontal Gyrus}    & OrbIF                          &          & {Superior Temporal Gyrus}            & {SupTemp}   \\
& {Orbital Part of the Medial Frontal Gyrus}      & {OrbMF}   &          & {Temporal Pole (Middle)}             & {MidTP}     \\
& {Orbital Part of the Middle Frontal Gyrus}      & {OrbMidF} &          & {Transverse Temporal Gyrus}          & {TransTemp} \\
& {Orbital Part of the Superior Frontal Gyrus}    & {OrbSF}   & \multicolumn{2}{l}{Cingulate Cortex}         &                                  \\
& Paracentral Lobule                              & PCL                            &          & {Anterior Cingulate Cortex}          & AntCng                           \\
& {Precentral Gyrus}                              & {PreC}    &          & {Middle Cingulate Cortex}            & MidCng                           \\
& {Rolandic Operculum}                            & RolOper                        &          & {Posterior Cingulate Cortex}         & PostCng                          \\
& {Superior Frontal Gyrus}                        & SupF                           & \multicolumn{2}{l}{Insular Cortex}           &                                  \\
& {Supplementary Motor Area}                      & SMA                            &          & Insula                                                    & Ins                              \\
& {Triangular Part of the Inferior Frontal Gyrus} & TriIFG                         & \multicolumn{2}{l}{White Matter}             &                                  \\
\multicolumn{2}{l}{Parietal Lobe}               &                                &          & Brainstem                                                 & {BS}        \\
& {Angular Gyrus}                                 & Ang                            &          & Corpus Callosum                                           & CC                               \\
& Inferior Parietal Lobule                        & {IPL}     &          & Internal Capsule                                          & IC                               \\
& {Postcentral Gyrus}                             & Pstcent                        &          & Periventricular White Matter                              & pWM                              \\
& Precuneus                                       & Precuneus                      & \multicolumn{2}{l}{Deep Gray Nuclei}         &                                  \\
& Superior Parietal Lobule                        & SPL                            &          & Amygdala                                                  & Amyg                             \\
& {Supramarginal Gyrus}                           & {SMG}     &          & {Caudate Nucleus}                    & Caud                             \\
\multicolumn{2}{l}{Occipital Lobe}              &                                &          & Lentiform                                                   & Lent                             \\
& {Calcarine Cortex}                              & Calc                           &          & Thalamus                                                  & Thal                             \\
& Cuneus                                          & Cuneus                         & \multicolumn{2}{l}{Cerebellum}               &                                  \\
& {Fusiform Gyrus}                                & Fusiform                       &          & Cerebellum                                                & Cb                               \\
& {Inferior Occipital Gyrus}                      & {InfOcc}  &          &                                                           &                                  \\
& {Lingual Gyrus}                                 & Ling                           &          &                                                           &                                  \\
& {Middle Occipital Gyrus}                        & {MidOcc}  &          &                                                           &                                  \\
& {Superior Occipital Gyrus}                      & {SupOcc}  &          &                              \\
 \thickhline
\end{tabular}
\end{table*}

\end{document}